\documentstyle[preprint,aps,epsf]{revtex}

\def\half{{\textstyle{1\over2}}}

\def\pmb#1{\setbox0=\hbox{$#1$}%
\kern-.025em\copy0\kern-\wd0
\kern.05em\copy0\kern-\wd0
\kern-.025em\raise.0433em\box0}

\def\beq{\begin{equation}}
\def\eeq{\end{equation}}

\begin{document}

\def\footnoterule{\hrule width \hsize}
\def\footstrut{\baselineskip 16pt}

\skip\footins = 14pt % was 12
\footskip     = 20pt % was 18
\footnotesep  = 12pt % was 10

\textwidth=6.5in
\hsize=6.5in
\oddsidemargin=0in
\evensidemargin=0in
\hoffset=0in

\textheight=9.5in
\vsize=9.5in
\topmargin=-.5in
\voffset=-.3in

\baselineskip=24pt plus .5pt

\title{% 
NON-YANG-MILLS 
GAUGE THEORIES
}

\footnotetext[1] {\baselineskip=16pt This work is supported in part by funds
provided by  the U.S.~Department of Energy (D.O.E.) under contract
\#DE-FC02-94ER40818. \hfil MIT-CTP-2628 \hfil  April 1997\break}

\author{R.~Jackiw\footnotemark[1]}

\address{Center for Theoretical Physics\\ Massachusetts Institute of
Technology\\ Cambridge, MA ~02139--4307}

\maketitle

\setcounter{page}{0}
\thispagestyle{empty}

\begin{abstract}%
\noindent%
Various gauge invariant but non-Yang-Mills dynamical models are
discussed: Pr\'ecis of Chern-Simons theory in $(2+1)$-dimensions and reduction
to $(1+1)$-dimensional $B$-$F$ theories; gauge theories for $(1+1)$-dimensional
gravity-matter interactions; parity and gauge invariant mass term in
$(2+1)$-dimensions.
\end{abstract}
%\vspace*{\fill}

\vskip 2in

\centerline{Non-Perturbative Quantum Field Physics, Pe\~niscola, Spain, June
1997}

\maketitle

\newpage

\section{Introduction}

The successes of present day theories for fundamental processes in Nature
offer persuasive evidence that forces between elementary particles obey the
principle of local gauge symmetry.  Even Einstein's gravity theory, which
thus far has not been incorporated into the quantal formalism that describes
all the other fundamental interactions, is seen to enjoy, at least at the
classical level, an invariance against local transformations, {\it viz.\/}
diffeomorphism invariance.

While the ``standard'' particle physics model realizes the gauge principle with
the Yang-Mills paradigm -- the non-Abelian generalization of Maxwell's
electrodynamics -- in the last decade we have come to appreciate that there
are other forms of dynamics that are gauge invariant, physically relevant,
but do not use the Yang-Mills structure.  These alternative
realizations of gauge invariance play only a small role in elementary
particle field theory, but they seem to have many applications to
phenomenological descriptions of various collective, emergent phenomena.  Also
they are mathematically fascinating.  

In my lectures here, I shall describe some recent work on non-Yang-Mills
gauge theories.  Let me begin by setting notation.  Vector gauge potentials
(connections) and vector gauge fields (curvatures) will be variously
presented in component notation ($A^a_\mu, F^a_{\mu \nu} \equiv \partial_\mu
A^a_\nu - \partial_\nu A^a_\mu + f_{bc} \!\!\! {\phantom A}^a \! A^b_\mu
A^c_\nu$), or by Lie-algebra valued relations ($A_\mu = A^a_\mu T_a, F_{\mu
\nu} = F^a_{\mu \nu} T_a = \partial_\mu A_\nu - \partial_\nu A_\mu + [A_\mu,
A_\nu]$).  The Lie-algebra generators satisfy $[T_a, T_b] = f_{ab}  \!\!\!
{\phantom T}^c  T_c$, thereby defining the structure constants.  We shall
assume that there exists an invariant, non-singular bi-linear $\langle T_a,
T_b \rangle =
\eta_{ab}$, with which group indices can be moved, so that $f_{abc} \propto
f_{ab}
\!\!\! {\phantom \eta}^d  \eta_{dc}$ is totally anti-symmetric.  Mostly
$\eta_{ab}$ will be the Killing-Cartan metric, but other structures can arise
with non-semi-simple groups.  Specifically for the $SU(N)$ fundamental
representation, the Pauli/Gell-Mann matrices are used: $T_a =
\lambda_a /2 i$, $\langle T_a, T_b\rangle = tr T_a T_b = -\half \delta_{ab}$. 
Finally, we shall also use form notation: $A = A_\mu dx^\mu, F = \half F_{\mu
\nu} dx^\mu dx^\nu = dA +A^2$, with no explicit indication of the outer
product.  In Minkowski space-time, our metric tensor is positive in its
time-time component.

\section{Pr\'ecis of the Chern-Simons Term in Three Dimensions}

The most important and popular non-Yang-Mills gauge structure is the
Chern-Simons term.  Physicists have been working with this quantity for
over a decade, so the subject is well-established, and I need not
discuss it here in detail.  Nevertheless a few introductory remarks will be
made, because transformations of the Chern-Simons term lead to the new models
that I wish to describe.

The Chern-Simons density made its first appearance in physics when it was
realized that the anomaly in the conservation law for the axial vector current
can be written as a divergence,
\begin{mathletters}
\begin{equation}
- \frac{1}{64 \pi^2} \epsilon^{\mu \nu \alpha \beta} F^a_{\mu \nu}
F^a_{\alpha \beta} = \partial_\mu K^\mu
\label{eq:1a}
\end{equation}  
\begin{equation}
K^\mu = - \frac{1}{16 \pi^2} \epsilon^{\mu \alpha \beta \gamma} \left\{
A_\alpha^a \partial_\beta A^a_\gamma + \frac{1}{3} f_{abc} A^a_\alpha
A_\beta^b A^c_\gamma \right\}
\label{eq:1b}
\end{equation}  
or in compact notation for 4-forms.
\begin{equation}
\frac{1}{8 \pi^2} \langle F, F\rangle = \frac{1}{8 \pi^2} d \langle A, dA +
\frac{2}{3}A^2\rangle
\label{eq:1c}
\end{equation}
\end{mathletters}%
Because $K^\mu$ contracts the Levi-Civita $\epsilon$-tensor, any one of its component
contains no fields in the direction of that component, and when the
coordinate of the selected direction is also suppressed, one naturally
arrives at a three-dimensional quantity.
\begin{mathletters}
\begin{eqnarray}
\Omega (A) &=& - \frac{1}{16 \pi^2} \epsilon^{\alpha \beta \gamma} \left\{
A_\alpha^a \partial_\beta A^a_\gamma + \frac{1}{3} f_{abc} A^a_\alpha
A_\beta^b A^c_\gamma \right\} \nonumber \\
 &=& - \frac{1}{32 \pi^2} \epsilon^{\alpha \beta \gamma} \left\{
A_\alpha^a F^a_{\beta \gamma} - \frac{1}{3} f_{abc}  A^a_\alpha
A_\beta^b A^c_\gamma \right\} \label{eq:2a} \\
   &=& - \frac{1}{16 \pi^2} \left\{ A_\alpha^a F^{\alpha a} - \frac{1}{6} 
\epsilon^{\alpha \beta \gamma} f_{abc} A^a_\alpha A_\beta^b A^c_\gamma \right\}
\label{eq:2b} \\
   &=& \frac{1}{8 \pi^2} \langle A, dA + \frac{2}{3}A^2\rangle \nonumber \\
   &=& \frac{1}{8 \pi^2} \langle A, F - \frac{1}{3}A^2\rangle \label{eq:2c} 
\end{eqnarray}
\end{mathletters}
In (\ref{eq:2b}) we have introduced the dual field.
\begin{equation}
F^{\alpha a} = \half \epsilon^{\alpha \beta \gamma} F^a_{\beta \gamma}
\label{eq:3}
\end{equation}
The 3-form $\Omega (A)$ is the Chern-Simons density, while its integral over
three-space is the Chern-Simons term $W(A)$.
\begin{equation}
W(A) = \int \Omega (A)
\label{eq:4}
\end{equation}
$W(A)$ possess the important property of being invariant against
infinitesimal gauge transformation, while changing under finite gauge
transformations by the integer winding number $n$ of the group element that
effects the transformation.
\begin{eqnarray}
\Omega (A^U) &=& \Omega (A) + \frac{1}{8 \pi^2} d \langle A, d UU^{-1}\rangle
-\frac{1}{48
\pi^2} \langle dUU^{-1}, dUU^{-1} dUU^{-1}\rangle \label{eq:5} \\
A^U &\equiv& U^{-1} AU + U^{-1} dU \label{eq:6} \\
W (A^U) &=& = W(A) +n \label{eq:7} \\
n &=&  -\frac{1}{48 \pi^2} \int \langle dUU^{-1}, dUU^{-1} dUU^{-1}\rangle
\label{eq:8}
\end{eqnarray}
We have assumed that no surface terms contribute: $\int d \langle A,
dUU^{-1}\rangle = 0$.

The discontinuous gauge variance (\ref{eq:7}) of $W(A)$ leads to the first
application of the Chern-Simons term in particle physics: the establishment
of the QCD vacuum angle.  Within a Hamiltonian, Schr\"odinger representation
approach to $(3+1)$-dimensional Yang-Mills quantum theory, states are
functionals of $A$, and in this fixed-time formalism the vector potential is
defined on three-space.  Gauss' law requires that these functionals be invariant
against infinitesimal gauge transformations, while finite gauge
transformations, which are symmetries of the theory, must leave states
invariant up to a phase.  So we immediately see that a physical state can
take the form
\begin{mathletters}
\begin{equation}
| \Psi\rangle = e^{i \theta W(A)} \Psi (A)
\label{eq:9a}
\end{equation}
where $\Psi(A)$ is invariant against {\bf all} gauge transformations. 
Consequently, in view of (\ref{eq:7}) a physical state responds to a gauge
transformation
$U$ as
\begin{equation}
\raise1pt\hbox{$| \Psi\rangle $}  
\lower3pt\hbox{${\longrightarrow \atop \raise2pt\hbox{$\scriptstyle U$}}$}
\, \raise3pt\hbox{$e^{i \theta n} |\Psi \rangle$}
\label{eq:9b}
\end{equation}
\end{mathletters}%
We see that the existence of the vacuum angle is here established without
reference to any instanton/tunnelling approximation.

[A further tantalizing fact is that $e^{\pm 8 \pi^2 W(A)}$ solves the
Yang-Mills functional Schr\"odinger equation with zero eigenvalue. 
Nevertheless this remarkable wave functional cannot represent a physical
state, because it grows uncontrollably for large $A$.]

Dynamical utilization of the Chern-Simons term came when it was suggested
that $W(A)$ can be a contribution to the action of a three-dimensional field
theory. \cite{ref:1,ref:2}  However with non-Abelian gauge groups, the
magnitude with which $W(A)$ enters must be quantized, in integer unites of $2
\pi$, so that the gauge variance of $W(A)$ produces a shift in the action of
$2 \pi \times$ integer, which would not be seen in the phase exponential of the
action.  (Sometimes it is claimed that this is also required for Abelian
groups, when the base manifold is topologically non-trivial.  But this is not
true -- in the Abelian case, a well-defined quantum theory can be defined for
arbitrary coupling strength, regardless of base-space topology.)

When $W(A)$ is added to the usual Yang-Mills term, the excitations of the
theory become massive, while retaining gauge invariance, but reflection
symmetry is lost.  One may couple further matter fields to $A_\mu$ in a gauge
invariant fashion.  For low-energy phenomenological applications to physical
systems confined on the plane, it makes sense to couple nonrelativistic matter. 
Moreover since the Yang-Mills term contains two derivatives and the
Chern-Simons term only one, the latter dominates the former at low energies,
and the Yang-Mills kinetic term may be dropped.  In this way one is led to the
interesting class of non Yang-Mills gauge theoretic  models described by
\begin{eqnarray}
I &=& 2\pi \kappa W(A) + \int \{i\psi^\ast D_t \psi - \frac{1}{2m}({\bf D}\psi)^\ast \cdot
({\bf D}\psi) - V (\rho)\} \nonumber \\
\rho & \equiv& \psi^\ast \psi \label{eq:10} 
\end{eqnarray}
where $\kappa$ is an integer (in the non-Abelian case), $(D_t,{\bf D})$ are
(temporal, spatial) gauge covariant derivatives, while $V(\rho)$ describes
matter self-interactions.  The matter action is Galileo invariant, the
Chern-Simons action is topological, {\it  i.e.\/} invariant against {\bf all}
coordinate transformations, so
$I$ is Galileo invariant.

Models belonging to the general class (\ref{eq:10}), with both Abelian and non-Abelian
gauge groups, have been widely discussed in the past, so I shall not describe that old
work here.\cite{ref:3,ref:4}  Rather I shall consider dimensional reduction of
(\ref{eq:10}) to
$(1+1)$ dimensions, and thereby expose some new non-Yang Mills, gauge theoretic
dynamical systems.

\section{Dimensional Reduction of Chern-Simons Theory=BF Theory}

One obvious reduction for the dynamics of (\ref{eq:10}) is to reduce in {\bf time}. This
corresponds to looking for static solutions to the Euler-Lagrange equations arising from
(\ref{eq:10}), and once again this subject is an old one, well reviewed in the existing
literature.\cite{ref:3,ref:4}  So I shall not dwell on it, beyond the reminder
that with an appropriate choice for $V$ one can apply the Bogomolny procedure
and replace the second-order Euler-Lagrange equations with coupled first order
equations, which in turn can be combined into the completely integrable Toda
equation (non-Abelian case) or Liouville equation (Abelian case), with
well-known soliton solutions.

\subsection{Reduction to non-Linear Schr\"odinger Equation}

Now I shall describe in detail a reduction to one {\bf spatial} dimension, which
results in an interesting reformulation of the non-linear Schr\"odinger equation.  On the
plane, with coordinates
$(x,y)$, we suppress all $y$-dependence and redefine $A_y$ as $B$.  Then, in
the Abelian case, the action (\ref{eq:10}) becomes
\begin{equation}
I = \int dt dx \left\{-\kappa BF + i\psi ^\ast D_t \psi - \frac{1}{2m}
|D\psi |^2  - \frac{1}{2m} B^2\rho - V(\rho)\right\}
\label{eq:11}
\end{equation}
The ``kinetic" gauge field term is the so-called ``$B$-$F$" expression where
$F=\half
\epsilon^{\mu\nu} F_{\mu\nu} = -\dot{a}-A^\prime_0$.  [We have
re-named $A_1$ as $-a$, and dot/dash refer to differentiation with respect to
time/space, {\it i.e.\/} $(t/x)$.  The covariant derivatives read $D_t\psi =
\dot{\psi}+iA_0\psi, D\psi=\psi^\prime-ia\psi$.  Also $\kappa$ has been
rescaled by $4 \pi$; recall that in the Abelian application, the Chern-Simons
coefficient is not quantized.]  Evidently the 2-dimensional $B$-$F$ quantity is
a dimensional reduction of the Chern-Simons expression.

Because (\ref{eq:11}) is first-order in time-derivatives, the action is already
in canonical form, and may be analyzed using the symplectic Hamiltonian
procedure.\cite{ref:5}  We present (\ref{eq:11}) as
\begin{eqnarray}
I & = & \int \left\{\kappa B \dot{a} + i\psi^\ast \dot{\psi} - A_0
\left(\kappa B^\prime + \rho\right) - \frac{1}{2m} |(\partial_x-ia)\psi|^2 -
\frac{1}{2m} B^2 \rho - V(\rho)\right\} \nonumber\\
& = & \int \left\{\kappa B\dot{a} + i\psi^\ast\dot{\psi}-A_0
\left(\kappa B^\prime+\rho\right) - \frac{1}{2m}
|(\partial_x-ia\pm B)\psi|^2 \mp\frac{1}{2m}  B^\prime\rho - V(\rho) \right\}
\nonumber \\
\label{eq:12}
\end{eqnarray}
$A_0$ is Lagrange multiplier, enforcing the Gauss law, which in this theory requires
\begin{mathletters}
\begin{equation}
B^\prime = - \frac{1}{\kappa} \rho 
\label{eq:13a}
\end{equation}
or equivalently
\begin{equation}
B(x) = -\frac{1}{2\kappa} \int d \tilde{x} \epsilon(x-\tilde{x})\rho(\tilde{x})
\label{eq:13b}
\end{equation}
\end{mathletters}%
[The Green's function, uniquely determined by parity invariance, is the
Heaviside $\pm 1$ step.]  Thus, after we eliminate $B$, (\ref{eq:12}) involves 
a spatially non-local Lagrangian.
\begin{mathletters}
\begin{eqnarray} 
L &=& - \half  \int dx d\tilde{x}
\dot{a}(x)\epsilon(x-\tilde{x})\rho(\tilde{x})+ \int dx i \psi^\ast\dot{\psi} 
\nonumber \\  
&& -  \frac{1}{2m} \int dx \bigg| \left(\partial_x - ia \mp
\frac{1}{2\kappa} \int d \tilde{x}
\epsilon(x-\tilde{x})\rho(\tilde{x})\right)\psi(x)\bigg|^2 \nonumber \\
&& + \int dx \left(\pm \frac{1}{2\kappa m}\rho^2-V(\rho)\right) 
\label{eq:14a}
\end{eqnarray}
The $a$ dependence is removed when $\psi(x)$ is replaced by $e^{\frac{i}{2} \int
d \tilde{x} \epsilon (x-\tilde{x}) a(\tilde{x})} \psi(x)$, leaving 
\begin{eqnarray} 
L &=&  \int dxi \psi^\ast\dot{\psi} -  \frac{1}{2m} \int dx \bigg|
\left(\partial_x \mp
\frac{1}{2\kappa} \int d\tilde{x}
\epsilon(x-\tilde{x})\rho(\tilde{x})\right)\psi(x)\bigg|^2
\nonumber \\ 
&& + \int dx \left(\pm \frac{1}{2\kappa m}\rho^2-V(\rho)\right) 
\label{eq:14b}
\label{eq:14}
\end{eqnarray}
\end{mathletters}%

Finally we choose $V(\rho)$ to be $\pm \frac{1}{2\kappa m}\rho^2$ [this is the same
choice that in $(2+1)$-dimensions leads to static first-order Bogomolny
equations] and our reduced Chern-Simons, $B$-$F$ theory is governed by the
Hamiltonian
\begin{equation}
H = \frac{1}{2m}\int dx \bigg| \left(\partial_x \mp \frac{1}{2\kappa} \int
 d\tilde{x} \epsilon(x-\tilde{x})\rho(\tilde{x})\right)\psi(x)\bigg|^2
\label{eq:15}
\end{equation}
which implies the first-order, Bogomolny equation
\begin{equation}
\psi'(x)\mp \frac{1}{2\kappa}\int d\tilde{x}
\epsilon(x-\tilde{x})\rho(\tilde{x})\psi(x)=0
\label{eq:16}
\end{equation}

On the other hand, we can recognize the dynamics described in (\ref{eq:15}) by
expanding the product.
\begin{eqnarray}
 H  = \frac{1}{2m} &&\int dx \left\{ |\psi'|^2 \pm \frac{1}{\kappa}
\rho^2 \right\}
\nonumber \\
+ \frac{1}{24m\kappa^2} &&\int dx d\tilde{x}
d \hat{x} \rho(x) \rho(\tilde{x}) \rho(\hat{x}) \left\{
\epsilon(x-\tilde{x}) \epsilon(x-\hat{x}) +
\epsilon(\tilde{x}-\hat{x})\epsilon(\tilde{x}-x) + \epsilon
(\hat{x}-x)\epsilon(\hat{x}-\tilde{x})\right\} \nonumber \\
{\phantom 1}&&
\label{eq:17}
\end{eqnarray}
The last term was symmetrized, leading to a sum of step function products,
which in fact equals to 1.  Consequently the last integral
is $\frac{1}{24m\kappa^2}N^3$, where $N=\int dx \rho(x)$, which is conserved in
the dynamics implied by (\ref{eq:17}).  Hence this term can be removed by
redefining
\begin{equation}
\psi \rightarrow e^{-i \frac{N^2t}{8m\kappa^2}} \psi
\label{eq:18}
\end{equation}
What is left is recognized as the Hamiltonian for the non-linear Schr\"odinger equation, with
equation of motion
\begin{eqnarray}
i\dot{\psi} & = & -\frac{1}{2m} \psi '' -\lambda \rho\psi \nonumber \\
\lambda & \equiv & \mp \frac{1}{m\kappa}
\label{eq:19}
\end{eqnarray}

The non-linear Schr\"odinger equation plays a cycle of interrelated roles in
mathematical physics.  Viewed as a non-linear, partial differential equation for the
function $\psi$, it is completely integrable, possessing a complete spectrum of
multi-soliton solutions, the simplest of these being the single soliton at
rest.  This requires
$\lambda > 0$, which is always achievable in our reduction by adjustment
of
$\kappa$.
\begin{equation}
\psi_s^{\rm rest} (t,x)= \pm e^{i\frac{\alpha^2}{2m}t} \frac{1}{\sqrt{\lambda
m}}\frac{\alpha}{\cosh
\alpha x}
\label{eq:20}
\end{equation}
($\alpha$ is an integration constant.)  Because of Galileo invariance, the solution may be
boosted with velocity $v$, yielding
\begin{equation}
\psi_s^{\rm moving} (t,x)= \pm e^{imvx} e^{it\left(\frac{\alpha^2}{2m} -
\frac{mv^2}{2}\right)} \frac{1}{\sqrt{\lambda m}} \frac{\alpha}{\cosh \alpha
(x-vt)}
\label{eq:21}
\end{equation}
The soliton solutions can be quantized by the well-known methods of soliton
quantization.  On the other hand, the non-relativistic field theory can be quantized at
fixed $N$, where it describes $N$ non-relativistic point particles with pair-wise
$\delta$-function interactions.  This quantal problem can also be solved
exactly, and the results agree with those of soliton quantization.  All these
properties are well-known, and will not be reviewed here.\cite{ref:6}

The present development demonstrates that this classical/quantal completely
integrable theory possesses a Bogomolny  formulation, which is obtained by
using two-dimensional $B$-$F$ gauge theory, which in turn descends from
three-dimensional Chern-Simons dynamics.  Indeed it is clear that (\ref{eq:20})
also solves the first-order equation (\ref{eq:16}) -- even the phase, which is
undetermined by (\ref{eq:16}), is consistent with (\ref{eq:18}).\cite{ref:7}

\subsection{Reduction to Modified non-Linear Schr\"odinger Equation}

While the previous development started with $B$-$F$ gauge theory, which
descended from a Chern-Simons model, and arrived at an interesting (first-order,
Bogomolny) formulation for the familiar non-linear Schr\"odinger equation, we
now further modify the gauge theory and obtain a novel, chiral, non-linear
Schr\"odinger equation.

Let us observe first that the above dynamics is non-trivial solely
because we have chosen $V$ to be non-vanishing.  Indeed with $V=0$ in
(\ref{eq:11}), (\ref{eq:12}) and (14), the same set of steps
(removing $B$ and $a$ from the theory) results in a free theory for the $\psi$
field.

To avoid triviality at $V=0$, we need to make the $B$ field dynamically active
by endowing it with a kinetic term.  Such a kinetic term could take the
Klein-Gordon form; however we prefer a simpler expression that describes a
``chiral'' Bose field, propagating only in one direction.  A Lagrange density
for such a field has been known for some time. \cite{ref:8}
\begin{equation}
{\cal L}_{\rm chiral} = \pm \dot{B} B' + v B' B'
\label{eq:22}
\end{equation}
Here $v$ is a velocity, and the consequent equation of motion arising from
${\cal L}_{\rm chiral}$ is solved by $B=B(x \mp vt)$ (with suitable boundary
conditions at infinity), describing propagation in one direction, with velocity
$\pm v$.  Note that $\dot{B} B'$ is {\bf not} invariant against a Galileo boost,
which is a symmetry of $B'B'$ and of (\ref{eq:11}), (\ref{eq:12}), (14):
performing a Galileo boost on $\dot{B} B'$ with velocity $\tilde{v}$ gives rise
to
$\tilde{v} B' B'$, effectively boosting the $v$ parameter in ${\cal L}_{\rm
chiral}$ by $\tilde{v}$. Consequently, one may drop the $v B' B'$
contribution, thereby selecting to work in a global ``rest frame.''  Boosting
a solution in this rest frame produces a solution to the theory with a $B' B'$
term.

In view of this discussion, we supplement the previous Lagrange density
(\ref{eq:11}), (\ref{eq:12}), (14) by $\pm \dot{B} B'$, set $V$ to
zero, and thereby replace (\ref{eq:12}) by
\begin{mathletters}
\begin{equation}
{\cal L} = - \kappa \dot{B} \left( a \mp \frac{1}{\kappa} B' \right) + i
\psi^\ast \dot{\psi} - A_0 (\kappa B' + \rho) - \frac{1}{2m} | (\partial_x -
ia) \psi |^2 - \frac{1}{2m} B^2 \rho
\label{eq:23a}
\end{equation}
After redefining $a$ as $a \pm \frac{1}{\kappa} B'$, this becomes equivalent to 
\begin{equation}
{\cal L} = \kappa B \dot{a} + i \psi^\ast \dot{\psi} - A_0 (\kappa B' +
\rho) - \frac{1}{2m} | (\partial_x -
ia \mp i \frac{1}{\kappa} B') \psi |^2 - \frac{1}{2m} B^2 \rho
\label{eq:23b}
\end{equation}
\end{mathletters}%
Now we proceed as before: solve Gauss' law as in (13), remove $a$ by
a phase-redefinition of $\psi$, drop the last term in (\ref{eq:23b}) by a
further phase redefinition as in (\ref{eq:18}).  We are then left with 
\begin{equation}
{\cal L} = i \psi^\ast \dot{\psi} - \frac{1}{2m} | (\partial_x 
\pm i \frac{1}{\kappa^2} \rho) \psi |^2 
\label{eq:24}
\end{equation}
It has been suggested that this theory may be relevant to modeling quantum
Hall edge states. \cite{ref:9}

The Euler-Lagrange equation that follows from
(\ref{eq:24}) reads
\begin{equation}
i \partial_t \psi = - \frac{1}{2m} \left(
\partial_x \pm i\frac{1}{\kappa^2} \rho \right)^2 \psi \pm 
\frac{1}{\kappa^2} j \psi
\label{eq:25}
\end{equation}
where the current density $j$
\begin{equation}
j = \frac{1}{m} {\rm Im} \psi^\ast \left( \partial_x \pm i
\frac{1}{\kappa^2} \rho \right) \psi
\label{eq:26}
\end{equation}
is linked to $\rho$ by the continuity equation.
\begin{equation}
\partial_t \rho + \partial_x j = 0
\label{eq:27}
\end{equation}
Next we redefine the $\psi$ field by
\begin{equation}
\psi (t,x) = e^{\mp \frac{i}{\kappa^2} \int^x dy \rho (t,y)} \Psi (t,x)
\label{eq:28}
\end{equation}
and see that the equations satisfied by $\Psi$ is
\begin{mathletters}
\begin{equation}
i \dot{\Psi} (t,x) \pm \frac{1}{\kappa^2} \int^x dy \dot{\rho} (t,y) \Psi
(t,x) = - \frac{1}{2m} \Psi'' (t,x) \pm \frac{1}{\kappa^2} j(t,x) \Psi (t,x) 
\label{eq:29}
\end{equation}
But the integral may be evaluated with the help of (\ref{eq:27}), so finally
we are left with \cite{ref:10}
\begin{equation}
i \dot{\Psi} = - \frac{1}{2m} \Psi'' \pm \frac{2}{\kappa^2} j \Psi
\label{eq:30}
\end{equation}
\end{mathletters}

This is a non-linear Schr\"odinger equation similar to (\ref{eq:19}) but with
the current density $j= \frac{1}{m} {\rm Im} \Psi^\ast \Psi'$ replacing the
charge density $\rho = \Psi^\ast \Psi$.  The equation is {\bf not} known to be
completely integrable but it does possess an interesting soliton solution, which
is readily found by setting the $x$ dependence of the phase of $\Psi$ to be
$e^{imvx}$.  Then $j= v\rho$, and our new equation (\ref{eq:30}) becomes the
usual non-linear Schr\"odinger equation (\ref{eq:19})
\begin{equation}
i \dot{\Psi} = - \frac{1}{2m} \Psi'' \pm \frac{2v}{\kappa^2} \rho \Psi
\label{eq:31}
\end{equation}
{\it i.e.\/} the non-linear coupling strength of (\ref{eq:19}) is 
\begin{equation}
\lambda = \mp \frac{2v}{\kappa^2} 
\label{eq:32}
\end{equation}

The $(\mp)$ sign is inherited from the ``chiral'' kinetic term, see
(\ref{eq:22}), (23); once a definite choice is made (say $+$),
positive $\lambda$, which is required for soliton binding, corresponds to
definite sign for $v$ (say positive); {\it i.e.\/} the soliton solving
(\ref{eq:31}) moves in only one direction.  Explicitly, with the above choice
of signs, the one-soliton solution reads
\begin{equation}
\Psi_s (t,x) = \pm e^{imvx} e^{it \left( \frac{\alpha^2}{2m} -
\frac{mv^2}{2} \right)} \frac{\kappa}{\sqrt{2mv}} \frac{\alpha}{\cosh
\alpha (x-vt)}
\label{eq:33}
\end{equation}
We see explicitly that $v$ must be positive; the soliton cannot be brought to
rest; Galileo invariance is lost.  

The characteristics of the solution are are follows
\begin{equation}
N_s =  \frac{\alpha \kappa^2}{mv}
\label{eq:34}
\end{equation}
The energy is obtained by integrating the Hamiltonian.
\begin{equation}
E = \int dx \frac{1}{2m} |\Psi'|^2
\label{eq:35}
\end{equation}
and on the solution (\ref{eq:33}), takes the value appropriate to a massive,
non-relativistic particle.
\begin{equation}
E_s = \half M_s v^2
\label{eq:36}
\end{equation}
where
\begin{equation}
M_s = mN_s (1 + \frac{1}{3 \kappa^4} N^2_s)
\label{eq:37}
\end{equation}
The conserved field momentum in this theory reads
\begin{equation}
P = \int dx (m j + \frac{1}{\kappa^2} \rho^2 )
\label{eq:38}
\end{equation}
and on the solution (\ref{eq:33}) its value again corresponds to that of a
massive, non-relativistic particle
\begin{equation}
P_s = M_s v
\label{eq:38_2}
\end{equation}
\begin{equation}
E_s = \frac{P_s^2}{2M_s}
\label{eq:39}
\end{equation}

As already remarked, the model is not Galileo invariant, but one can verify
that it is scale invariant.  Indeed one can show that the above kinematical
relations are a consequence of scale invariance.\cite{ref:10}

The soliton solution (\ref{eq:33}) can be quantized; also the quantal many
body problem, which is implied by (\ref{eq:24}), can be analyzed.  Because the
system does not appear integrable, exact results are unavailable, but one
verifies that at weak coupling, the two methods of quantization (soliton, many
body) produce identical results.\cite{ref:10,ref:11}

\section{More $B$-$F$ Theories}

$B$-$F$ theories can of course be extended to non-Abelian groups, with $B$
transforming as an adjoint vector, just as $F$, so that their inner product is
a group scalar.  (More generally, one can even dispense with the group metric,
by positing that $B$ transforms in the co-adjoint representation.) 
Two-dimensional Yang-Mills theory is a $B$-$F$ theory: since the Yang-Mills
action may be written as $I_{YM} = \frac{1}{2} \int d^2x F^a F^a$, where $F^a =
\frac{1}{2} \epsilon^{\mu \nu} F_{\mu\nu}^a$, equivalent dynamics is governed by
\begin{equation}
I = \int d^2x (B^a F^a - \half B^a B^a)
\label{eq:40}
\end{equation}
Moreover, since the $B$-$F$ contribution is a world scalar -- invariant against
{\bf all} coordinate transformations -- general coordinate invariance is broken
only by the second term, which however is invariant against area-preserving
coordinate transformation.  Hence the latter are seen as symmetries of
two-dimensional Yang-Mills theory, and this observation aids greatly in
unraveling that model on spaces with non-trivial topology.\cite{ref:12}  I do
not pursue this topic here, but turn to another role for $B$-$F$ theories:
gauge theoretic formulations of two-dimensional gravity.

\subsection{Two-Dimensional Gravity}

In order to have a gravity theory in two dimensions, where the Einstein tensor
vanishes identically and the Einstein-Hilbert action is a surface term,
therefore not generating Euler-Lagrange equations of motion, one introduces a
further, world scalar variable, these days called the dilaton $\eta$.  A class
of possible actions is 
\begin{equation}
I_{\rm 2-d~gravity} = \int d^2x \sqrt{-g} \Big(\eta R - V (\eta)\Big)
\label{eq:41}
\end{equation}
where different theories are selected by choosing various forms for $V$.  Two
especially interesting choices are $V (\eta) = \Lambda \eta$ and $V (\eta) =
\Lambda$, where $\Lambda$ is a (cosmological) constant.  The former is the
first-such model that was proposed in 1984\cite{ref:13}; the second is the
popular, string-inspired CGHS model.\cite{ref:14}

It now happens that precisely these two models can be formulated as
$B$-$F$ gauge theories, the former based on the $SO(2,1)$ de Sitter or anti-de
Sitter groups,\cite{ref:15} while the latter on the centrally extended
Poincar\'e group
$ISO(1,1)$.\cite{ref:16}  These constructions, which I described in my last
visit to a Spanish Summer School (Salamanca, 1992), are two-dimensional analogs
to the construction of three-dimensional Einstein theory as a Chern-Simons gauge
theory -- a discovery by the Spanish physicist Ana Ach\'ucarro, among
others.\cite{ref:17}

Let me quickly review how this is done.  Rather than using metric variables,
we introduce the {\it Zweibein} $e_\mu^{A}$ and the spin-connection
$\omega_\mu^{AB}=\epsilon^{AB} \omega_\mu$, which are viewed as independent
variables -- the relation between them emerges as an equation of motion.
[Capital Roman letters refer to the flat two-dimensional tangent space, with
metric tensor $\delta_{AB}$, $g_{\mu\nu} = e_\mu^{A} e_\nu^{B}
\delta_{AB}$.]  We view the spin connection as a vector potential
associated with the Lorentz rotation generator $J$ on the $(1+1)$-dimensional
Minkowski tangent space, while the {\it Zweibeine} are vector potentials
associated with translations $P_{A}$.  Next an algebra is postulated: the
commutator with
$J$ is conventional
\begin{equation}
[P_{A}, J] = \epsilon_{A} \!\!\! {\phantom P}^{B} P_{B}
\label{eq:42}
\end{equation}
The commutor of the translations takes different forms for $SO(2,1)$ and for
centrally extended $ISO(1,1)$,
\begin{eqnarray}
[P_{A}, P_{B}] &=& \epsilon_{AB} \Lambda J \qquad SO(2,1)
 \\ \label{eq:43}
[P_{A}, P_{B}] &=& \epsilon_{AB} \Lambda I \qquad ISO(1,1)
\label{eq:44}
\end{eqnarray}
Here $I$ is a central element, commuting with $P_{A}$ and $J$. 
Consequently in the extended $ISO(1,1)$ case, the group is enlarged from three
parameters to four, since $I$ is taken to be an additional (commuting)
generator, and a further vector potential $a_\mu$ is associated with it. 
(Henceforth we redefine the generators so that $\Lambda$ is scaled to unity.)

The potentials are collected into a group-valued connection
\begin{equation}
A_\mu = A_\mu^a T_a = e_\mu^A P_{A} + \omega_\mu J + a_\mu I
\label{eq:45}
\end{equation}
where the last term is present only in the centrally extended $ISO(1,1)$
case.  The field strength is computed in the usual manner
\begin{equation}
\epsilon_{\mu\nu} F= F_{\mu\nu} = F^a_{\mu\nu} T_a = \partial_\mu A_\nu
- \partial_\nu A_\mu + [A_\mu, A_\nu]
\label{eq:46}
\end{equation}
with the commutator evaluated from (\ref{eq:42}) - (\ref{eq:44}).  One finds
that the component of $F_{\mu\nu}$ along the translation direction is the
gravitational torsion, while that along the rotation direction is the
gravitational curvature.  

The gauge transformation properties are as expected: an infinitesimal gauge
parameter $\Theta$ is constructed in the Lie algebra, {\it i.e.\/} with an
expansion similar to (\ref{eq:45}); then $\delta A_\mu = D_\mu \Theta =
\partial_\mu \Theta + [A_\mu, \Theta], \delta  F = [F, \Theta]$, and from
these one can read off the transformation properties for the component
fields.

Finally to form the action, a set of Lagrange multiplier fields is introduced;
they transform in the (co-)adjoint representation and allow construction of
the scalar $B$-$F$ quantity ($B$ replaced by $\eta$)
\begin{equation}
I_{\rm 2-d~gravity} = \int d^2x \eta_a F^a
\label{eq:47}
\end{equation}
One readily verifies that the equations of motion that follow from
(\ref{eq:47}) coincide with those of (\ref{eq:41}) in the $SO(2,1)$ and
extended $ISO(1,1)$ cases.

While the geometric formulation (\ref{eq:41}) is equivalent to the gauge-group
formulation (\ref{eq:47}), the latter is much more readily analyzed, by
exploiting gauge invariance.  The equations of motion that follow from
(\ref{eq:47}) are
\begin{mathletters}
\begin{eqnarray}
F^a &=& 0
 \\ \label{eq:48a}
(D_\mu \eta) _a &=& \partial_\mu \eta_a + f_{ab} \!\!\! {\phantom A}^c \!
A^b_\mu \eta_c = 0
\label{eq:48b}
\end{eqnarray}
\end{mathletters}%
Classical solution is straightforward.  Eq.~(\ref{eq:48a}) requires $A$ to be
a pure gauge, and we may pass to the gauge $A=0$ (assuming that there is no
obstruction).  Then (48) states that in the chosen gauge $\eta_a$ is
constant.  Of course in this gauge the geometry is lost -- vanishing $A$ means
that the {\it Zweibeine} and spin-connection vanish.  But we can now return to
a non-trivial gauge $A=U^{-1} d U$, which leads to non-vanishing geometric
quantities.  On the other hand, an invariant must be constructed
solely from $\eta_a$, since $F^a$ vanishes.  To form $\eta_a \eta^{ab} \eta_b$
we need a group metric, and with $SO(2,1)$ the obvious expression is
the indefinite, diagonal Killing-Cartan metric.  The
Poincar\'e group, not being semi-simple, does not possess a non-singular
invariant metric, but its central extension does: one verifies that $\eta_a
\eta^{ab}
\eta_b = \eta_A
\delta^{AB} \eta_B - 2 \eta_2 \eta_3$ is indeed invariant (we label the four
indices $a: 0,1,2,3$, where $A: 0,1$ refers to the two-dimensional Minkowski
tangent space).

The quantum theory is also readily analyzed with the help of the gauge
formalism.

Upon writing the action (\ref{eq:47}) in canonical, first order form
\begin{equation}
I_{\rm 2-d~gravity} = \int d^2x \Big( \eta_a \dot{A}^a_1 + A^a_0 (D_1 \eta)_a
\Big) \label{eq:49}
\end{equation}
we recognize that $A^a_1$ and $\eta_a$ are the canonically conjugate
coordinates and momenta, respectively, and the only dynamical
equation is Gauss' law, which requires physical states to be
annihilated by $(D_1 \eta)_a$.  It is convenient to analyze this requirement
in the field theoretic, Schr\"odinger-``momentum'' representation, where
states are functionals of the canonical momentum, here $\eta_a, \Phi
(\eta)$.  Evidently these functionals must satisfy
\begin{equation}
\left(\eta'_a + i f_{ab} \!\! {\phantom \eta}^c \eta_c \frac{\delta}{\delta
\eta_b}\right) \Phi(\eta) =0
\label{eq:50}
\end{equation}
Solution is immediate.  We first observe that $\Phi$ has support only on
constant $\eta_a \eta^a$, as is exemplified by contracting (\ref{eq:50})
with
$\eta^a$, and using the anti-symmetry of $f_{abc}$ to drop the functional
derivative.  So we write
\begin{equation}
\eta = g^{-1} K g 
\label{eq:51}
\end{equation}
where $g$ is a group element and $K$ is constant and it follows that a
solution to (\ref{eq:50}) is 
\begin{eqnarray}
\Phi (\eta) &=& e^{iS(\eta)}
\nonumber \\ 
S(\eta) &=& - \int dx \langle K, g' g^{-1} \rangle \nonumber \\
&=& - \int dx \langle \eta, g^{-1} g' \rangle
\label{eq:52}
\end{eqnarray}
with $g$ related to $\eta$ by (\ref{eq:51}).

The above structure (\ref{eq:52}) has another role in mathematical physics,
quite distinct from the role in which we encounter it here as the phase of a
wave functional.  Observe that $S$ in (\ref{eq:52}) is given by an integral
of the 1-form $\langle K, dg g^{-1} \rangle$, which one may take as a
canonical 1-form for a Lagrangian with dynamical variables $g$ depending on
``time.''  It is then further true that the symplectic 2-form, $d \langle K,
dg g^{-1}
\rangle = \langle K, dg g^{-1}  dg g^{-1}\rangle$ defines Poisson brackets
and that the brackets of the quantities $Q^a = (g^{-1} Kg)^a$ reproduce the Lie
algebra of the relevant group.  This 2-form is associated with the names
Kirillov and Kostant.\cite{ref:18} (One recognized here the development that was
previously described as occurring in connection with the Chern-Simons term: an
expression with interesting gauge transformation properties arises first in
physics as the phase of a wave functional; subsequently it acquires its own
dynamical role in a lower-dimensional theory.)

\medskip

\centerline{ASIDE ON GAUGE THEORETIC WAVE FUNCTIONALS IN THE MOMENTUM
REPRESENTATION}

\medskip

Let us observe that the wave functional in (\ref{eq:52}) is {\bf not} gauge
invariant.  A gauge transformation on $\eta \to \eta^U = U^{-1} \eta U$ is
effected, according to (\ref{eq:51}), by $g \to gU$.  Hence we see in
(\ref{eq:52}) that
\begin{mathletters}
\begin{eqnarray}
\Phi (\eta^U) &=& e^{-i \Omega (\eta, U)} \Phi (\eta)
\label{eq:53a} \\ 
\Omega (\eta, U)  &=& \int \langle \eta, U' U^{-1} \rangle \ \ .
\label{eq:53b}
\end{eqnarray}
\end{mathletters}%
This feature is a universal property of gauge theoretic wave functionals in
the momentum representation -- a little-known fact that deserves elaboration.

We consider a typical gauge theory, in any number of dimensions, with a
conventional, non-Chern-Simons gauge Gauss law, which ensures that wave
functionals in the coordinate representation, {\it i.e.\/} depending on $A$,
are gauge invariant.  (We ignore the vacuum angle that may arise when
topologically non-trivial gauge transformations are considered.)  Note that the
present discussion does {\bf not} apply in the presence of a Chern-Simons term,
because then Gauss' law becomes unconventional and further effects are
present.  The momentum representation -- we call it the ``$E$'' representation,
because (for conventional, non-Chern-Simons gauge theories) the
$E^i$ field is conjugate to $A_i$ -- can be related to the coordinate
representation by a (functional) Fourier transform.
\begin{equation}
\Phi (E) = \int {\cal D} A \left(\exp -i \int \langle E^i, A_i \rangle \right)
\Psi (A)
\label{eq:54}
\end{equation}
Now the following sequence of equations holds.
\begin{eqnarray}
\Phi (E) &=& \int {\cal D} A \left(\exp -i \int \langle E, A \rangle \right)
\Psi (U^{-1} AU + U^{-1} \partial U)
\nonumber \\ 
  &=& \int {\cal D} A \left(\exp -i \int \langle E^U, A \rangle \right)
\Psi  (A + U^{-1} \partial U) \nonumber \\
  &=& \exp i \int \langle E, \partial UU^{-1} \rangle \int {\cal D} A
\left( \exp -i \int \langle E^U, A \rangle \right) \Psi  (A)
\nonumber \\
 &=& \exp i \Omega (E,U) \Phi (E^U)
\label{eq:55}
\end{eqnarray}
The first equality is true because $\Psi (A)$ is gauge invariant.  In the next
equality we have changed integration variables: $A \to U \, AU^{-1}$; this has
unit Jacobian, and affects the phase by replacing $E$ with its gauge transform
$E^U = U^{-1} EU$.  In the next step, $A$ is shifted: $A \to A - U^{-1}
\partial U$; this produces the phase $\Omega (E,U)$ seen in the last equality.
\begin{equation}
\Omega (E,U) =  \int d^d r E^i_a (\partial_i UU^{-1})^a
\label{eq:56}
\end{equation}
Thus from (\ref{eq:55}), it follows that physical wave functionals in the
``E'' representation are not gauge invariant.  Rather, after a gauge
transformation they acquire the phase $\Omega (E,U)$, 
\begin{equation}
\Phi (E^U) = e^{-i \Omega (E,U)} \Phi (E)
\label{eq:57}
\end{equation}
which is recognized to be a 1-cocycle {\it i.e.\/} $\Omega (E,U)$ satisfies
\begin{equation}
\Omega (E,U_1U_2) = \Omega (E^{U_1}, U_2) + \Omega (E,U_1)
\label{eq:58}
\end{equation}
as is required by (\ref{eq:56}) when two gauge transformations are composed.

We conclude therefore that physical functionals in the ``$E$''
representation, which are annihilated by the Gauss law generator $G_a$, obey
(\ref{eq:57}).  In one spatial dimension with $\eta$ replacing $E^i$, we
regain (53). 

\subsection{Incorporating Matter in Gauge Theories of Gravity}

While the gauge theoretical formulation of gravity in two dimensions (and also
in three) proceeds smoothly,\cite{ref:15,ref:16,ref:17} incorporating matter
poses new problems, because even when the {\it Zweibein}/spin connection are
employed in the matter Lagrangian, the gauge symmetry is not apparent.

Here I shall discuss the simplest case -- a point-particle coupled to gravity;
similar ideas apply when matter fields are coupled to gravity.

An action for a point particle of mass $m$ can be given in first-order form as
\begin{equation}
I = \int d\tau \Big( p_A e^A_\mu \dot{x}^\mu + \half N (p^2 -m^2) \Big)
\label{eq:59}
\end{equation}
Upon varying and eliminating $p_A= -e^A_\mu \dot{x}^\mu/N$ and $N=
\sqrt{\dot{x}^\mu g_{\mu \nu} \dot{x}^\nu}/m$ one recognizes the above
as the familiar second-order action $-m \int d\tau \sqrt{\dot{x}^\mu g_{\mu \nu}
\dot{x}^\nu}$.  Here $x^\mu (\tau)$ is the particle trajectory, on which the
gravitational quantities $g_{\mu\nu}$ and $e^A_\mu$ depend, while the
auxiliary quantities $p_A$ and $N$ depend on the parameter $\tau$, which can
be reparametrized at will.

Expression (\ref{eq:59}) is diffeomorphism invariant.  For its response to
gauge transformations, we adopt the CGHS model of gravity theory, in the
extended $ISO(1,1)$ gauge group formulation.  According to the general
discussion given previously, when the infinitesimal gauge function is taken as 
\begin{equation}
\Theta = \theta^A P_A + \alpha J + \beta I
\label{eq:60}
\end{equation}
the components of the gauge theoretic connection transform as 
\begin{eqnarray}
\delta e^A_\mu &=& - \alpha \epsilon^A \!\! {\phantom \epsilon}_B \,\, e^B_\mu +
\epsilon^A \!\! {\phantom \epsilon}_B \,\,  \theta^B
\omega_\mu + \partial_\mu \theta^A \nonumber \\
\delta \omega_\mu &=& \partial_\mu \alpha \nonumber \\
\delta a_\mu &=& e^A_\mu \epsilon_{AB} \theta^B + \partial_\mu \beta
\label{eq:61}
\end{eqnarray}
The natural gauge transformation law for the matter variables in (\ref{eq:59})
is that $x^\mu$ and $N$ are scalars, while $p_A$ responds by
\begin{equation}
\delta p_A = - \alpha \epsilon_A \!\! {\phantom \epsilon}^B
\,\, p_B
\label{eq:62}
\end{equation}
But then (\ref{eq:59}) is not invariant against local translations, generated
by $\theta^A$.

To remedy this situation we proceed as follows.  A new variable is introduced
$q^A$, called the {\bf Poincar\'e} coordinate, with infinitesimal gauge
transformation law
\begin{equation}
\delta q^A = - \alpha \epsilon^A \!\! {\phantom \epsilon}_B
\,\, q^B - \theta^A
\label{eq:63}
\end{equation}
{\it i.e.\/} under Lorentz transformations $q^A$ rotates in the usual way, but
is shifted by translations.  As a consequence, by a translational gauge
transformation, $q^A$ may be always set to zero.

A gauge invariant action is now constructed by replacing $e^A_\mu \dot{x}^\mu$
in (\ref{eq:59}) by $(D_\tau q)^A$, where
\begin{equation}
(D_\tau q)^A = \dot{q}^A + \epsilon^A \!\! {\phantom \epsilon}_B
\,\, q^B \omega_\mu \dot{x}^\mu + e^A_\mu \dot{x}^\mu
\label{eq:64}
\end{equation}
\begin{equation}
I_{\rm invariant} = \int d \tau \Big( p_A (D_\tau q)^A + \half N(p^2 -m^2)
\Big)
\label{eq:65}
\end{equation}
Invariance is established once it is verified from (\ref{eq:61}),
(\ref{eq:62}) and (\ref{eq:64}) that 
\begin{equation}
\delta (D_\tau q)^A  = - \alpha \epsilon^A \!\! {\phantom \epsilon}_B
\,\, (D_\tau q)^B
\label{eq:66}
\end{equation}
On the other hand, dynamics has not been changed because the Poincar\'e
coordinate can be set to zero by a gauge transformation and the action
(\ref{eq:59}) is regained from  (\ref{eq:65}).

In a sense this is a Higgs-like mechanism, with $q^A$ playing the role of a
``Goldstone'' field, which is needed for a manifestly symmetric formulation. 
In the ``unitary'' gauge the Goldstone field is absent, only physical degrees
of freedom are visible, but the symmetry is lost.\cite{ref:20}

A manifestly covariant group theoretical formalism is available, and one can
also accommodate matter fields by introducing a Poincar\'e field.  Within this
formalism, CGHS-matter quantum field theory has been analyzed completely, with
interesting results.  The most noteworthy of these shows that the
constraints, which are present in the theory, acquire quantum mechanical
obstructions (anomalies) whose form depends on the ordering prescription used
in defining the theory.  These anomalies can be removed by further adjustment
of the theory, and a ``physical'' spectrum of states is displayed.\cite{ref:21}

I shall not pursue this topic here further, beyond remarking that once the
route to a successful analysis is found within the gauge theoretical
formalism, it is then possible to identify analogous paths in the geometric
formulation of the theory, as well.\cite{ref:22}

%%section 5 goes here
\section{Parity-Even Mass for Three-Dimensional Gauge Fields}

I remarked previously that the Chern-Simons expression, when added to the
three-dimensional Yang-Mills action, renders the fields massive, while
preserving gauge invariance.  However, parity symmetry is lost.

A trivial way of maintaining parity with this mass generation is through the
doublet mechanism.  Consider a pair of identical Yang-Mills actions, each
supplemented with their own Chern-Simons term, which enters with opposite
signs.  The parity transformation is defined to include field exchange
accompanying coordinate reflection, and this is a symmetry of the doubled
theory.

But is there a way of maintaining reflection symmetry without introducing
parity doublets?  Here I shall show how this can be done, and I shall use
various ideas that I have already discussed.  But first, a bit of motivation.

Three dimensional gauge theories possess theoretical/mathematical interest, but
they merit study because they describe (1) kinematical processes that are
confined to a plane when external structures (magnetic fields, cosmic strings)
perpendicular to the plane are present, and (2) static properties of
$(3+1)$-dimensional systems in equilibrium with a high-temperature heat bath. 
An important issue is whether the apparently massless gauge theory possesses a
mass gap.  The suggestion that indeed it does gains support from the
observation that the gauge coupling constant squared carries dimension of mass,
thereby providing a natural mass-scale (as in the two-dimensional Schwinger
model).  Also without a mass gap, the perturbative  expansion is infrared
divergent, so if the theory is to have a perturbative definition, infrared
divergences must be screened, thereby providing evidence for magnetic screening
in the four-dimensional gauge theory at high temperature.

But in spite of the above indications, a compelling theoretical derivation of
the desired result is not yet available, even though many approaches have been
tried.

Here I shall not address the dynamical question of how such a mass gap can be
generated.  Rather I present a phenomenological construction:  I offer a theory
for massive vector fields, which is gauge invariant and parity
preserving.\cite{ref:23}

Consider the Lagrange density
\begin{equation}
{\cal L} = \ \mbox{tr}\left(F^\mu F_\mu + G^\mu G_\mu - 2m F^\mu
\Phi_\mu\right)
\label{eq:67}
\end{equation}
The first term is the usual Yang-Mills expression, written in terms of the dual
field
$F^\mu\equiv \frac{1}{2} \epsilon^{\mu\alpha\beta} F_{\alpha\beta}$; the second
describes a charged vector field $\Phi_{\mu}$ in the same adjoint
representation as the gauge potential and interacting with it.
\begin{equation}
G^\mu = \half \epsilon^{\mu\alpha\beta} G_{\alpha\beta}, \;
G_{\alpha\beta} = D_\alpha\Phi_\beta - D_\beta \Phi_\alpha \ \ , \qquad 
D_\alpha =
\partial_\alpha + [A_\alpha,{\phantom{A_\alpha}}]
\label{eq:68}
\end{equation}
Finally the last contribution carries the mass scale $m$ and involves a mixed
Chern-Simons-like structure,
$$F^\mu\Phi_\mu=\half \epsilon^{\mu\alpha\beta}F_{\alpha\beta}\Phi_\mu$$
Obviously the theory is gauge invariant.  

The equations of motion
\begin{mathletters}
\begin{equation}
\epsilon_{\mu\alpha\beta} D^\alpha F^\beta - m G_\mu +
\epsilon_{\mu\alpha\beta}[G^\alpha,\Phi^\beta] = 0 
\label{eq:69a}
\end{equation}
\begin{equation}
\epsilon_{\mu\alpha\beta} D^\alpha G^\beta - m F_\mu = 0 
\label{eq:69b}
\end{equation}
\end{mathletters}
may be combined into a second-order equation
\begin{equation}
D^2 G_\mu - D^\nu D_\mu G_\nu + m^2 G_\mu -
m\epsilon_{\mu\alpha\beta}[G^\alpha,\Phi^\beta]=0
\label{eq:70}
\end{equation}
The linear part of (\ref{eq:70}) shows the $G^\mu$ is massive, and then
(\ref{eq:69b}) shows that $F^\mu$ is massive as well.  By positing that
$\Phi_\mu$ carries odd parity, we ensure that reflection symmetry is maintained.

The theory possesses an interesting symmetry structure.  In addition to being
invariant against gauge transformations
\begin{equation}
\delta_1 A_\mu = D_\mu \theta, \quad \delta_1\Phi_\mu = [\Phi_\mu,\theta]
\label{eq:71}
\end{equation}
there is a another transformation
\begin{equation}
\delta_2 A_\mu = 0 \quad \delta_2\Phi_\mu = D_\mu \chi
\label{eq:72}
\end{equation}
which obviously does not affect the Yang-Mills term, and changes the
interaction/mixing term by total derivative, because $F^\mu$ obeys the Bianchi
identity
$D_\mu F^\mu = 0$.  However, the kinetic term for the $\Phi$ fields is not
invariant
\begin{equation}
\delta_2 G^\mu = [F^\mu,\chi]
\label{eq:73}
\end{equation}

To gain further insight, let us record the Lie algebra of the generators for
the symmetry group as
\begin{equation}
[Q_a,Q_b] = f_{ab} \!\!\! {\phantom f}^c  Q_c
\label{eq:74}
\end{equation}
and recall that the vector potential $A^a_\mu$ is the connection associated
with this group.

What group theoretical role can we assign to $\Phi^a_\mu$?  Note that this
field has as many components as $A^a_\mu$.  Let us therefore postulate an
Abelian group with as many parameters as in (\ref{eq:74}) and generators $P_a$,
satisfying
\begin{equation}
[P_a,P_b]=0
\label{eq:75}
\end{equation}
We shall consider the vector fields $\Phi^a_\mu$ to be connections in this
Abelian group.  Moreover, let us further postulate
\begin{equation}
[Q_a,P_b] = f_{ab} \!\!\! {\phantom f}^c   P_c
\label{eq:76}
\end{equation}

With these definitions we can unify much of our formalism.  Define the
connection on both groups as
\begin{equation}
{\cal A}_\mu = A_\mu^a Q_a + \Phi^a_\mu P_a
\label{eq:77}
\end{equation}
The curvature ${\cal F}_{\mu\nu} = \partial_\mu {\cal A}_\nu -
\partial_\nu{\cal A}_\mu + [{\cal A}_\mu, {\cal A}_\nu]$ is evaluated from
(\ref{eq:74})--(\ref{eq:77}), and nicely decomposes into $F^a_{\mu\nu}$ and
$G^a_{\mu\nu}$
\begin{equation}
{\cal F}_{\mu\nu} = F^a_{\mu\nu} Q_a + G^a_{\mu\nu} P_a
\label{eq:78}
\end{equation}
Also the transformations (\ref{eq:71}), (\ref{eq:72}) can be collected together
by defining the infinitesimal gauge parameter
\begin{equation}
\Theta = \theta^\alpha Q_a + \chi^a P_a
\label{eq:79}
\end{equation}
and recognizing that (\ref{eq:71}), (\ref{eq:72}) are equivalent to
\begin{equation}
\delta {\cal A}_\mu = {\cal D}_\mu \Theta = \partial_\mu \Theta + [{\cal A}_\mu,
\Theta]
\label{eq:80}
\end{equation}

But in spite of the above unification of formalism, the action associated with
(\ref{eq:67}) remains non-invariant, since $G^2$ is not invariant.  [This is
because
$Q_a Q^a + P_a P^a$ is {\bf not} an invariant of the algebra
(\ref{eq:74})--(\ref{eq:76}).] However, this defect can be fixed, in a manner
similar to the construction of a gauge invariant gravity-matter interaction in
two dimensions, which I described earlier.  To this end, we introduce an
additional scalar field multiplet $\rho^a$, which transforms under
(\ref{eq:71}) as an adjoint vector
\begin{equation}
\delta_1 \rho^a = f_{bc} \!\!\! {\phantom f}^a   \rho_b\theta_c
\label{eq:81}
\end{equation}
while the transformation (\ref{eq:72}) effects a shift
\begin{equation}
\delta_2\rho^a = -\chi^a
\label{eq:82}
\end{equation}
Also, in ${\cal L}$ we replace $G^\mu$ by $G^\mu + [F^\mu,\rho]$, which is
invariant: $\delta_2(G^\mu + [F^\mu,\rho])=[F^\mu,\chi]+[F^\mu,-\chi]=0$.

Thus we have constructed an invariant, non-Yang-Mills theory governed by the
Lagrange density
\begin{equation}
{\cal L}_\rho = - \frac{1}{4} F^a_{\mu\nu} F^{\mu\nu a} - \frac{1}{4}
(G^a_{\mu\nu}+f_{abc} F^b_{\mu\nu}\rho^c)(G^{\mu\nu a}+f_{abc} F^{\mu\nu b}
\rho^c) + \frac{m}{2}
\epsilon^{\alpha\beta\gamma} F^a_{\alpha\beta}\Phi_\gamma^a
\label{eq:83}
\end{equation}
Moreover, the dynamics is the same as that of ${\cal L}$ in (\ref{eq:67}),
because with the gauge transformation (\ref{eq:82}) one can set $\rho$ to zero,
thereby reducing
${\cal L}_\rho$ to ${\cal L}$, in this ``unitary" gauge.\cite{ref:24}

It is worth commenting on the unconventional aspects of this realization for
gauge invariant, but non-Yang-Mills dynamics.  Usually when considering gauge
fields and vector fields that are ``charged'' with respect to the gauge
group, one includes couplings beyond the minimal ones, and embeds everything
(gauge fields, vector fields) in a larger non-Abelian gauge group.  [For
example, electrically charged vector fields and Maxwell gauge fields are endowed
with non-minimal interactions and combined into and $SU(2)$ non-Abelian group,
with the ``third'' direction being electromagnetism and the other ``two''
referring to the charged degrees of freedom.]  On the other hand, we have
combined our group degrees of freedom into a semi-direct product structure,
(\ref{eq:74})-(\ref{eq:76}), and invariance is achieved without non-minimal
coupling, other than introducing the ``Goldstone'' field $\rho^a$, which
disappears in the ``unitary'' gauge.

Although formal quantization of the model can be carried out,\cite{ref:23}
developing a perturbative calculational method requires further analysis.  The
point is that the quadratic portion of the kinetic term for the $\Phi_\mu$ field
does {\bf not} define a propagator, because the derivative operator is
transverse and has no inverse.  On the other hand, it seems impossible to 
resolve this problem by a gauge fixing term -- as is done for the
$A_\mu$ kinetic term. One possibility is to use a background $A_\mu$ field to
define the
$\Phi_\mu$ propagator.  Indeed the existence of the ``Goldstone'' field, which
shifts by a constant under a symmetry transformation, hints at some kind of
symmetry breaking.

This model deserves further study, so its properties can be completely
understood.

\end{document}